\documentclass[
    12pt,
    a4paper
    ,preprint
    ,pra
    ,aps
    ,oneside
    ,nobibnotes
    ,floatfix
    ,showpacs
]{revtex4-1}

\usepackage{ifpdf}

\usepackage{amsmath,amssymb,bm}
\usepackage{txfonts}
\usepackage[T1]{fontenc}
\usepackage[utf8]{inputenc}
\usepackage[english]{babel}
\usepackage{graphicx}\graphicspath{{Graph/}{../Graph/}}

    \DeclareMathOperator{\e}{e}

    \let\Re=\undefined
    \newcommand{\Re}{\mathrm{Re}}
    \let\Im=\undefined
    \newcommand{\Im}{\mathrm{Im}}

    \newcommand{\parder}[2]{\frac{\partial #1}{\partial #2}}

    \DeclareMathOperator{\dif}{d\!}
    \newcommand{\der}[2]{\frac{\dif{#1}}{\dif{#2}}}
    \newcommand{\tder}[2]{{\dif{#1}}/{\dif{#2}}}
    
    \DeclareMathOperator{\erfi}{erfi}

    \renewcommand{\vec}[1]{\mathbf{#1}}

    \usepackage{xcolor}
    \definecolor{darkblue}{cmyk}{1.00, 0.50, 0.00, 0.40}

\begin{document}
\title{Diffraction of surface wave on conducting rectangular wedge}

\author{Igor A. Kotelnikov, Vasily V. Gerasimov, Boris A. Knyazev}
    \affiliation{Novosibirsk State University, Pirogova Street 2, Novosibirsk, Russia}
    \affiliation{Budker Institute of Nuclear Physics SB RAS, Lavrentyeva Av. 11, Novosibirsk, Russia}

\begin{abstract}
    Diffraction of a surface wave on a rectangular wedge with impedance faces is studied using the Sommerfeld-Malyuzhinets technique. An analog of Landau's bypass rule in the theory of plasma waves is introduced for selection of a correct branch of the Sommerfeld integral, and the exact solution is given in terms of imaginary error function. The formula derived is valid both in the near-field and far-wave zones. It is shown that a diffracted surface wave is completely scattered into freely propagating electromagnetic waves and neither reflected nor transmitted surface waves are generated in case of bare metals which have positive real part of surface impedance. The scattered waves propagate predominantly at a grazing angle along the direction of propagation of the incident surface wave and mainly in the upper hemisphere regarding the wedge face. The profile of radiated intensity is nonmonotonic and does not resemble the surface wave profile which exponentially evanesces with the distance from the wedge face.

    Comparison with experiments carried out in the terahertz spectral range at Novosibirsk free electron laser has shown a good agreement of the theory and the experiments.
\end{abstract}


\maketitle




\section{Introduction}
\label{s1}

Plasmonics is now a very rapidly developing field of activity, and advancement to the mid- and far-infrared (terahertz) ranges of frequencies is one of the main streams in photonics \cite{Maier2007, Stanley2012NP_6_409}. Our interest in this subject has arisen from discussions of the results of experiments on propagation of Surface Plasmon Polaritons (SPPs) along gold-ZnS-air interfaces \cite{Gerasimov+2012b}, which were carried out at Novosibirsk Free Electron Laser facility \cite{Knyazev+2010MST_21_054017} in the terahertz spectral range. For SPPs not to be disturbed by any material probe, plasmon-polariton characteristics were studied indirectly via sensing of the electromagnetic field in the space behind the tail facet of samples. In the experiments, the profile of radiated intensity is non-monotonic and does not resemble a surface wave  profile which exponentially evanesces with the distance from the wedge face. A maximum of the radiation intensity is observed at some distance from the surface plane whereas a maximum of the surface wave field should be located at the wedge surface.
These discrepancies had initiated search for theoretical explanation which has led us to the theory developed by G.\,D.~Malyuzhinets in the 1950s.

The Sommerfeld-Malyuzhinets theory is known to provide a powerful method for exact solution to specific optical problems \cite{BorovikovKinber1994, Osipov+1999WM_29_313,
Babich+2008}.
In particular, G.D. Malyuzhinets solved the problem of diffraction of a surface wave by an impedance wedge, considering it as a special case of a plane wave propagation at the Brewster angle \cite{Malyuzhinets1959SovPhysDokl_3_752}. He derived some general relations but he did not investigate the properties of integrand functions near the saddle points which is necessary for computation of the  integrals. A few books and review papers \cite{BorovikovKinber1994, Osipov+1999WM_29_313,
Babich+2008} of varying elaboration provide a systematic introduction to the theory by G.D. Malyuzhinets. Some earlier approaches to the problem of diffraction by wedges and screens are reviewed in \cite{Zon2007JOSA(B)_24_1960,Zon2007JOptA_9_476}. We would like to add a few references to this list \cite{Trenev1958RadioElektron_3_27, Trenev1958RadioElektron_3_163, Starovoitova+1962RadioElektron_2_250, Ivanov1970146, Ivanov1971349, Lyalinov1996RadioScience, Lyalinov1999275, Gautesen2005239, Lyalinov+2006WM_44_21, Norris+1999WM_30_69, Shimoda+2001IEICETranslectr_7_853}.


%
%
%

V.~Zon in Ref.~\cite{Zon2007JOSA(B)_24_1960,Zon2007JOptA_9_476} applied the
Sommerfeld-Malyuzhinets technique  to the diffraction of a surface wave on conducting wedge with small surface impedance but she did not succeed in obtaining the final solution. We do that in this paper. We also conclude that her calculations of the reflection and transmission coefficients for the surface wave are incorrect. To obtain correct results we introduce a sort of Landau's bypass rule known in the theory of plasma waves \cite{Shafranov1967}.

In Section \ref{s2}, we briefly recall the properties of surface waves to a minimal extent necessary for understanding of the subsequent calculations and justify the use of the Leontovich boundary conditions at the wedge faces; in particular, we give (without derivation) a formula for the surface impedance of a metal substrate coated with a thin dielectric film.  The foundations of the Sommerfeld-Malyuzhinets technique are expounded in Section \ref{s3}. The diffracted field is calculated in Sections \ref{s4}, \ref{s5}, and \ref{s6}. A brief description of the experimental techniques and comparison of experimental results with theoretical predictions are given in Section \ref{s7}. Section \ref{s9} concludes the paper and summarizes our results.

\section{Surface Plasmons on a Plane}
\label{s2}


Let a metal with the complex permittivity $\varepsilon$ occupy the lower half-space $y<0$. Since a TE wave (the electric field is transverse to the plane of incidence) has no surface branch, we consider a TM wave in the upper half-space $y>0$, localized near the metal-air interface and propagating in the $x$-direction:
    \begin{equation}
    \label{2:1}
    \begin{split}
    \vec{E}(\vec{r},t) &= \vec{E}_{0}
    \exp\left[- \varkappa y + i(kx - \omega t)\right]
    ,
    \qquad
    \vec{E}_{0} = \{E_{0x},E_{0y}, 0\},
    \\
    \vec{B}(\vec{r},t) &= \vec{B}_{0}
    \exp\left[- \varkappa y + i(kx - \omega t)\right]
    ,
    \qquad
    \vec{B}_{0} = \{0, 0, B_{0z}\}
    .
    \end{split}
    \end{equation}
It is characterized by the circular frequency  $\omega$, wavenumber $k$, and attenuation constant $\varkappa$ with a positive real part, $\Re \varkappa>0$. Similar fields (with a negative attenuation constant, $\Re \varkappa'<0$) could be written for the lower half-space but they are needed only for derivation of the dispersion law for the surface wave:
    \begin{equation}
    \label{2:2}
    k = 
        k_{0} \sqrt{\frac{\varepsilon}{1+\varepsilon}}
    .
    \end{equation}
Here
    \begin{equation*}
    k_{0}=\omega /c
    ,
    \qquad
    \varkappa^2=k^2-k_{0}^2
    ,
    \end{equation*}
the Gaussian system of units is used, and the derivation can be found elsewhere (see \cite{Raether1988, Klimov2009eng}). Eq.~\eqref{2:1} can be interpreted as an evanescent plane wave (also known as the Zenneck wave \cite{Ufimtsev+1998,OHara+_33_245(2012)}) propagating at a complex angle $\chi$ such that
    \begin{equation}
    \label{2:3}
    k_x = k_{0} \cos\chi = k
    ,
    \qquad
    k_y = -k_{0} \sin\chi = i\varkappa
    ,
    \qquad
    \varkappa = i k_{0} \sin\chi
    .
    \end{equation}
Using dispersion relation of Eq.~\eqref{2:2} yields
    \begin{equation}
    \label{2:5}
    \cos\chi = \sqrt{\frac{\varepsilon}{1+\varepsilon}}
    ,
    \qquad
    \sin\chi = \sqrt{\frac{1}{1+\varepsilon}}
    .
    \end{equation}
In a fictitious case of real $\varepsilon$, the surface wave exists provided that $\varepsilon<-1$. For complex values of $\varepsilon$ there is always a solution with
    \(
    \Im \chi <0
    \) 
which exponentially decreases with the distance from the boundary of the metal. However, speaking about surface waves it makes sense only if the damping length $1/\Im k_{x}$ is sufficiently long, i.e. $|\Im k_{x}|\ll |\Re k_{x}|$. This condition is certainly satisfied if
    \(
    |\varepsilon| \gg 1.
    \)
Then
    \(
        k_{x} \approx k_{0}\left(
            1- {1}/{2\varepsilon}
        \right)
    \)
and the imaginary part of $k_{x}$ is automatically small no matter how large or small the imaginary part of $\varepsilon$ is.

Keeping the terahertz radiation at Novosibirsk Free Electron Laser \cite{Knyazev+2010MST_21_054017} in mind, we will take as an example the value of permittivity
    \begin{equation*}
    \varepsilon = -103260 + i\,310810 
    \end{equation*}
for gold at a frequency corresponding  to the wavelength
$\lambda = 140\,\text{microns}$
\cite{Ordal+1985AO_24_4493}. Then
    \begin{gather*}
    \chi=0.001022 - i\,0.001417 
    ,
    \\
    k_{x}=(44880 + i\, 0.065)\,\text{m}^{-1} 
    ,
    \\
    \varkappa = (64 + i\, 46)\,\text{m}^{-1} 
    .
    \end{gather*}


The components of the electric and magnetic field at the metal surface are related through the boundary condition, which can be derived from the Maxwell equation
    \begin{equation*}
    \nabla\times\vec{B} = \frac{1}{c}\parder{\vec{E}}{t}
    .
    \end{equation*}
Its $x$-component
    \begin{equation*}
    -\varkappa B_{0z} = -i k_{0} E_{0x}
    \end{equation*}
together with Eq.~\eqref{2:3} yields the Leontovich boundary condition
    \(
    E_{0x} = \sin\chi B_{0z}
    .
    \)
In the general case, the Leontovich boundary condition relates the tangential component of the electric field $\vec{E}_{\tau}$ with that of the magnetic field. It is usually written in the form
    \(
    \vec{E}_{\tau} = \xi \left[\vec{n}\times \vec{B}\right]
    ,
    \)
where the unit vector $\vec{n}$ is directed along the outward normal to the surface of the metal, and the parameter $\xi$ is called (dimensionless) surface impedance.

In Leontovich's theory, $\xi$ is assumed to be a function of frequency, which depends only on the conductor material but neither on the incident angle nor the type of incident wave. Using Fresnel's formulae one can readily check that
    \[
        \xi = \sqrt{\frac{1}{\varepsilon}-\frac{\sin^2{\theta_{0}}}{\varepsilon^2}}
        ,
    \]
for a plane TM wave impinging against the metal surface under the angle $\theta_{0}$ to the normal. A propagating TE wave is characterized by a different surface impedance
    \[
        \xi = \frac{1}{\sqrt{\varepsilon-\sin^2\theta_{0}}}
        ,
    \]
and
    \begin{equation}
    \label{2:11}
        \xi = \sin\chi = \sqrt{\frac{1}{1+\varepsilon}}
    \end{equation}
for the surface wave \eqref{2:1}. In any case,
    \[
    \label{2:18}
        \xi \simeq 1/\sqrt{\varepsilon}
        ,
    \]
provided that inequation $|\varepsilon| \gg 1$ holds. We see that for large $|\varepsilon|$ the surface impedance $\xi$ is approximately independent on parameters of the waves, which justifies using the Leontovich boundary condition on a metal surface instead of looking for complete solution for wave propagation inside a metal.

Below we will write the Leontovich boundary condition in the following form
    \begin{equation}
    \label{2:19}
    \vec{E}_{\tau} = \sin\chi \left[\vec{n}\times \vec{B}\right]
    ,
    \end{equation}
assuming that the formal parameter $\chi$ is related with the surface impedance by Eq.~\eqref{2:11}. As  shown above, $\chi$ can be considered as a complex angle of propagation of surface wave. For the above cited parameters of golden surface,
    \begin{equation}
    \xi = 0.00102-i\,0.00142 
    .
    \end{equation}

Coating of metallic surfaces strongly affects the magnitude of $\xi$. For a metal coated with a thin film with a small width $d\ll 2\pi/k_0$ and permittivity $\epsilon_{d}\gg 1$, we derived the following expression for the surface impedance:
    \begin{equation}
    \label{2:22}
    \xi
    \simeq
        \frac{1
        }{
             \sqrt{\varepsilon}
        }
        - i
        \left(
            \frac{\varepsilon_{d}-1}{\varepsilon_{d}}
        \right)
        k_{0} d
    .
    \end{equation}
Properties of the film becomes dominating for $d\gtrsim 1/k_{0}\sqrt{\varepsilon}$. For example, ZnS film with $d=0.75\mu\text{m}$ and
    \begin{equation*}
    \varepsilon_{d}=8.7 + i\,0.059
    \end{equation*}
over a gold substrate has the impedance
    \begin{equation}
    \label{2:25}
    \xi \simeq
    0.00106 - i\,0.0312 
    \end{equation}
with an imaginary part more than $20$ times larger than that of pure gold.

\section{Sommerfeld-Malyuzhinets theory}
\label{s3}

\begin{figure}
  \includegraphics[width=0.5\textwidth]{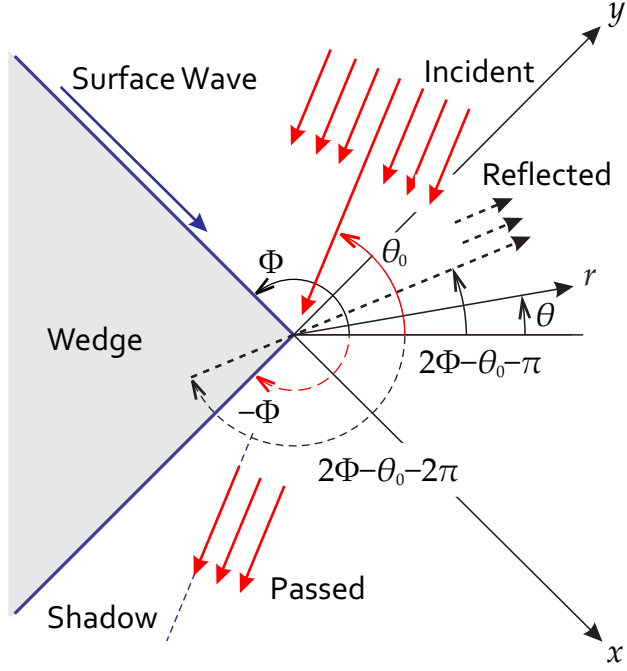}
  \caption{
    (color online)
    The geometry of the problem in cylindrical coordinates.
    The angles $\theta$ and $\Phi$ are measured from the bisector of the wedge.
  }
  \label{Fig:1}
\end{figure}

To begin with, let consider a plane wave of unit amplitude incident on a wedge at the angle $\theta=\theta_0>0$, as shown in Fig.~\ref{Fig:1}. In polar coordinates $r$ and $\theta$ perpendicular to the edge of the wedge, the wave is written as
    \begin{equation}
    \label{3:1}
      B_\text{inc} = \exp\left[
        - i k_{0} r \cos(\theta-\theta_{0})
      \right]
      .
    \end{equation}
Let the wedge occupy the region $|\theta|\geq \Phi$. Within the framework of geometrical optics, the upper face of the wedge reflects the incident wave at the angle $\theta=2\Phi-\theta_{0}-\pi$. One more $\pi$ should be subtracted from this value for  taking the reversal of the direction of propagation into account, thus the field of the reflected wave is derived from Eq.~\eqref{3:1} with the substitution $\theta_{0}\to 2\Phi-\theta_{0}-2\pi\to 2\Phi-\theta_{0}$:
    \begin{equation}
    \label{3:2}
      B_\text{refl} \propto \exp\left[
        - i k_{0} r \cos(\theta+\theta_{0}-2\Phi)
      \right]
      .
    \end{equation}

Below we restrict ourselves to the case of TM waves with the transversal component of magnetic field $B_{z}\equiv B$. The reason for such choice is that a surface wave may have no TE polarization. In the vacuum region, $|\theta|<\Phi$, the TM wave obeys the Helmholtz equation
\begin{equation}
    \label{3:3}
    \nabla^2 B + k_{0}^2 B =0
    .
\end{equation}
The non-zero components of the electric field in the TM wave,
\begin{equation}
    \label{3:4}
    E_{r} = \frac{i}{k_{0}r}\parder{B}{\theta}
    ,
    \qquad
    E_{\theta} = - \frac{i}{k_{0}}\parder{B}{r}
    ,
\end{equation}
are expressed through the derivatives of $B$. The Leontovich boundary conditions in
Eq.~\eqref{2:19} with the surface impedance of Eq.~\eqref{2:11}
on the conducting faces of  the wedge at $\theta=\pm\Phi$ now take the following form
\begin{gather}
    \label{3:5}
    \frac{1}{r}\parder{B}{\theta}
    =
    \pm ik_{0} \sin\chi_{\pm}\,B
    ,
\end{gather}
where the $+$ and $-$ signs label the quantities related to the faces of the
wedge located at the angle $\theta = +\Phi$ and $\theta = -\Phi$, respectively.

\begin{figure}
  \includegraphics[width=0.5\textwidth]{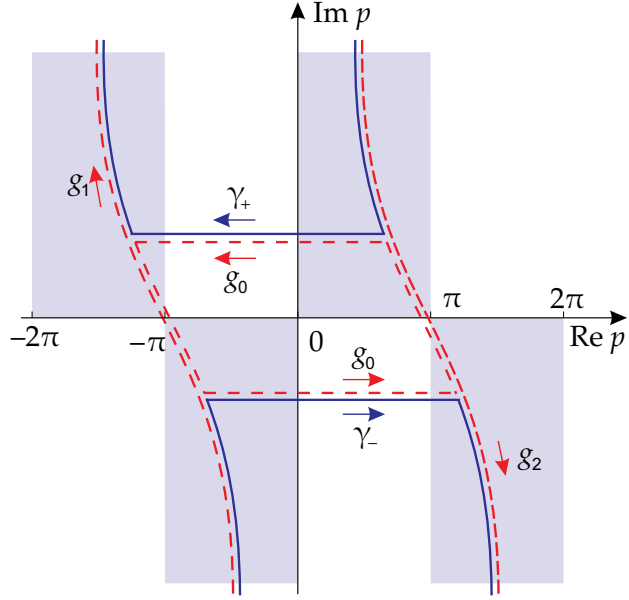}
  \caption{
    (color online)
    Plane of the complex angle $p$ with the Sommerfeld double loops $\gamma_{+}\cup\gamma_{-}$ (solid line) and the steepest descent paths $g_{0}\cup g_{1}\cup g_{2}$ (dashed line).
  }
  \label{Fig:2}
\end{figure}

Heuristic considerations enable construction of a solution to Eq.~\eqref{3:3} with given boundary conditions \cite{Babich+2008}. According to Sommerfeld \cite{Sommerfeld1896MathAnn_47_317}, the solution is sought in the form of a superposition of plane waves
\begin{equation}
    \label{3:7}
    B(r,\theta)
    =
    \frac{1}{2\pi i}
    \int\limits_{\gamma}
    \exp\left(-ik_{0}r\cos p\right)
    s(\theta - p)
    \dif p
    ,
\end{equation}
where the integration is carried out over the contour $\gamma$ in the complex $p$-plane, $k_{0}r>0$ and the kernel $s(\theta-p)=s(\theta-p, \theta_{0},\Phi)$ is analytic outside the real axis. The shadowed areas in Figure \ref{Fig:2} indicate the regions of the complex $p$-plane where $\Im\cos p < 0$ and the exponent of Eq.~\eqref{3:7} tends to zero as $kr\to \infty$.

Following Sommerfeld, we choose $\gamma$ in the form of two loops. Then, the integral in Eq.~\eqref{3:7} is a solution to Helmholtz's equation, provided that the two loops, $\gamma_{+}$ and $\gamma_{-}$, are shifted upwards and downwards correspondingly by the distance $|\Im(\theta-p)|>|\Im(\chi)|$; thus any singularities (see below) of the function $s(\theta-p)$ are located between these loops within the strip $-|\Im(\chi)|<\Im(\theta-p)<|\Im(\chi)|$. Choosing the integration contour in the form of two symmetrical loops guaranties that the  asymptote of the integral in Eq.~\eqref{3:7} at $k_{0}r\to \infty$ contains no converging cylindrical waves except for the incident wave of Eq.~\eqref{3:1}, and the incident wave is introduced into $s$ through given a pole at $p=\theta_0$ as explained in \cite{Babich+2008}.

Let now write functional equations for $s$. For the Leontovich boundary conditions of Eq.~\eqref{3:3}, we have
\begin{equation}
    \int\limits_{\gamma}
    \exp\left(-ik_{0}r\cos p\right)
    \left[
        \frac{1}{r}\,
        s'(\pm\Phi - p)
        \pm
        ik_{0} \sin\chi_{\pm}\,s(\pm\Phi-p)
    \right]
    \dif p
    =
    0
    ,
\end{equation}
where
$s'=\tder{s}{p}=-\tder{s}{\theta}$.
Integrating the first term by parts yields the following equations
\begin{equation}
    \label{3:10}
    \int\limits_{\gamma}
    \exp\left(-ik_{0}r\cos p\right)
    \left[
        -\sin p
        \pm \sin\chi_{\pm}
    \right]
    s(\pm\Phi-p)
    \dif p
    =
    0
    .
\end{equation}
Since the contour $\gamma=\gamma_{+}\cup\gamma_{-}$ consists of two loops situated symmetrically to the point $p = 0$, these equations are true if the kernel $s$ is an even function of $p$. This leads us to the Malyuzhinets functional equations \cite{Malyuzhinets1959SovPhysDokl_3_752}:
    \begin{equation}
    \label{3:11}
    \begin{split}
    \left[
         \sin p + \sin\chi_{+}
    \right] s( p + \Phi)
    &=
    \left[
        -\sin p + \sin\chi_{+}
    \right] s(-p + \Phi )
    ,
    \\
    \left[
         \sin p - \sin\chi_{-}
    \right] s( p - \Phi)
    &=
    \left[
        -\sin p - \sin\chi_{-}
    \right] s(-p - \Phi )
    .
    \end{split}
    \end{equation}
With rather non-trivial calculations one can verify \cite{Babich+2008} that the function
\begin{gather}
    \label{3:12}
    \Psi_{0}(u) =
        \Psi\left(u,\chi_{+}\right)
        \Psi\left(u-2\Phi,\chi_{-}\right)
    ,
    \\
\intertext{where}
    \label{3:13}
    \Psi\left(v,\chi\right) =
        \psi_{\Phi}\left(v+\Phi+\tfrac{1}{2}\pi-\chi\right)
        \psi_{\Phi}\left(v+\Phi-\tfrac{1}{2}\pi+\chi\right)
    ,\\
    \label{3:14}
    \psi_{\Phi}\left(w\right) =
        \exp\left[
            -\frac{1}{2}
            \int_{0}^{\infty}
            \frac{
                \cosh(w \eta) - 1
            }{
                \eta
                \cosh\left(
                    \tfrac{1}{2}\pi \eta
                \right)
                \sinh\left(
                    2\Phi \eta
                \right)
            }
        \dif\eta
        \right]
        ,
\end{gather}
supplies a particular solution for $s(u)$ to the functional equations \eqref{3:11}.
The Malyuzhinets function $\psi_{\Phi}$ is regular in the strip $|\Re (p)| < \tfrac{1}{2}\pi + 2\Phi$, where the integral in Eq.~\eqref{3:14} is even and satisfies the following functional relation
    \begin{gather}
    \label{3:15a}
    \psi_{\Phi}\left(w+2\Phi\right)
    /
    \psi_{\Phi}\left(w-2\Phi\right)
    =
    \cot\left(\tfrac{1}{2}w+\tfrac{1}{4}\pi\right)
    .
    \end{gather}
With Eq.~\eqref{3:15a} true this function extends beyond the indicated  band.

For $\Phi = m\pi /4n$, where $m/n$ is an irreducible rational number, the derivative $\tder{\ln\psi_{\phi}}{p}$ of $\ln\psi_{\phi}$ can be written as a finite sum of trigonometric functions \cite{Malyuzhinets1959SovPhysDokl_3_752}. In particular,
\begin{equation}
    \label{3:16}
    \psi_{3\pi/4}\left(w\right)
    =
    \frac{4}{3}
    \cos\left(\frac{w-\pi }{6}\right)
    \cos\left(\frac{w+\pi }{6}\right)
    \sec\left(\frac{w}{6}\right)
    =
    \frac{4}{3} \cos\left(\frac{w}{6}\right)
    -\frac{1}{3} \sec\left(\frac{w}{6}\right)
\end{equation}
for $\Phi = \tfrac{3}{4}\pi$, which corresponds to a right angle wedge.

Looking for a general solution in the form
\begin{equation}
    \label{3:17}
    s(u) = \sigma(u)\Psi_{0}(u)/\Psi_{0}(\theta_{0})
    ,
\end{equation}
we deduce from Eqs.~\eqref{3:11} that the function $\sigma(u)$ obeys the following equations
\begin{equation}
    \label{3:18}
    \begin{split}
    \sigma( p + \Phi)
    &=
    \sigma(-p + \Phi )
    ,
    \\
    \sigma( p - \Phi)
    &=
    \sigma(-p - \Phi )
    .
    \end{split}
    \end{equation}
For description of an incident wave with unit amplitude, $\sigma(\theta-p)$ should have a simple pole at $p=\theta_{0}$ with unit residue. By virtue of Eqs.~\eqref{3:18}, $\sigma$
is symmetric about points $p = \pm\Phi$, and therefore its system of poles
should satisfy the same symmetry relationships. If the pole $p = \theta_{0}$ is successively reflected about points $p = \Phi$ and $p = -\Phi$, one obtains for $s(\theta-p)$ a lattice of poles with the period $4\Phi$ at $p = \theta_0 + 4n\Phi$ ($n = 0, \pm 1, \pm 2$, etc.) and a similar grid at points $p = 2\Phi - \theta_{0} + 4n\Phi$. It is possible to guess a function with these properties. It  was found by G.\,D.~Malyuzhinets \cite{Malyuzhinets1958SovPhysDokl_3_52}:
\begin{equation*}
    \sigma(u)
    =
    \frac{
        \mu
        \cos(\mu\theta_{0})
    }{
        \sin(\mu u)
        -
        \sin(\mu\theta_{0})
    }
    ,
\end{equation*}
where $\mu=\pi/2\Phi$.  For $\Phi=\frac{3}{4}\pi$
\begin{equation}
    \label{3:20}
    \sigma(u)
    =
    \frac{
        \frac{2}{3}
        \cos\left(\frac{2\theta_{0}}{3}\right)
    }{
        \sin\left(\frac{2u}{3}\right)
        -
        \sin\left(\frac{2\theta_{0}}{3}\right)
    }
    .
\end{equation}



\section{Approximation of geometrical optics}
\label{s4}

At a first glance, the Sommerfeld-Malyuzhinets theory might be sought of as a purely mathematical trick. However, it indeed describes a real physical phenomenon. To prove that, in this Section we show how an approximation of geometrical optics can be retrieved from this theory which provides an \emph{exact} solution of the problem of vector field diffraction, although for a very limited set of problems related to the diffraction of electromagnetic waves on a conducting wedge. To avoid possible misunderstanding regarding the role of geometrical optics in the theory to be presented in the next three sections, it is worthy to emphasize that the field which will be derived in the current section is merely a part of a complete solution. From mathematical point of view, it represents a contribution of the poles of the integrand core $s(\theta-p)$ in Eq.~\eqref{3:7} whereas a complete solution \eqref{5:19} includes also a contribution of the saddle points
$p=\pm\pi$. The separation of the contribution of the poles and saddle points can be approximately made for the distances $r$ from the wedge top larger than the wavelength, $k_{0}r\gg 1$.

Following G.D. Malyuzhinets \cite{Malyuzhinets1959SovPhysDokl_3_752}, we replace the contour $\gamma=\gamma_{+}\cup\gamma_{-}$ with the 3-loop contour $g_0\cup g_1\cup g_2$ shown in Fig.~\ref{Fig:2} with the dashed line, in order to compute the Sommerfeld integral in Eq.~\eqref{3:7}:
\begin{equation}
    \label{4:1}
    B(r,\theta)
    =
    \frac{1}{2\pi i}
    \int\limits_{g_0\cup g_1 \cup g_2}
    \exp\left(-ik_{0}r\cos p\right) s(\theta-p) \dif p
    .
\end{equation}
The contours $g_0$, $g_1$ and $g_2$ are obtained by deformation of the contours $\gamma_{+}$ and $\gamma_{-}$ into the steepest descent paths through the saddle points $p=-\pi$ and $p=\pi$ of the exponent $\exp(-ik_{0}r\cos p)$ in the integrand. The new contour $g_0$ is formed from the middle parts of the contours $\gamma_{+}$ and $\gamma_{-}$, whereas the contours $g_{1}$ and $g_{2}$ are composed by remaining end parts of $\gamma_{+}$ and $\gamma_{-}$. As will be shown in Section \ref{s5}, the contribution of the saddle point describes the diffracted field at large distances (in the wave zone).

The integral over the contour $g_0$ is reduced to the sum of the residues in the poles of the integrand encircled by $g_0$ within the strip
\begin{equation}
    \label{4:2}
    - \pi < \Re (p) < \pi
    .
\end{equation}
These are the poles to be crossed during the deformation of the original contour of integration $\gamma_{1}\cup\gamma_{2}$ to the steepest descent path contour $g_{1}\cup g_{2}$. The number of such poles depends on the values of $\Phi$, $\theta$, and $\theta_0$.

Let consider first the poles of the function $\sigma$ in Eq.~\eqref{3:17}. There are an infinite number of such poles located at the points
    \begin{equation}
    \label{4:3}
    p_n =
    \theta - (-1)^n \theta_{0} - 2 n\Phi,
    \qquad
    n =0, \pm1,\dots
    .
    \end{equation}
However only three of them, namely
\begin{equation}
    \label{4:5}
    p_0 =
    \theta - \theta_{0}
    ,    \qquad
    p_1 =
    \theta + \theta_{0} - 2\Phi
    ,    \qquad
    p_{-1} =
    \theta + \theta_{0} + 2\Phi
    ,
\end{equation}
have chance to fall into the region \eqref{4:2} if $\Phi>\tfrac{1}{2}\pi$.

The residue at the pole $p_{0}$ stands for the wave $\exp[-ik_{0}r\cos(\theta-\theta_0)]$, which is readily recognized as the incident wave. It enters into the result of integration if $-\pi < \theta-\theta_{0} < \pi$. Values of $\theta$ outside the interval $-\pi + \theta_{0} < \theta < \pi + \theta_{0}$ and within the interval $-\Phi < \theta <\Phi$ represent the regions of geometrical shadow for the incident wave.

The pole $p_{1}$ gives rise to the wave $\exp[-ik_{0}r\cos(\theta+\theta_0-2\Phi)] =\exp[ik_{0}r\cos(\theta+\theta_0-2\Phi+\pi)]$, reflected from the upper face $\theta=\Phi$ of the wedge (see Fig.~\ref{Fig:1}). It enters into the result of integration if $-\pi < \theta+\theta_{0}-2\Phi < \pi$, i.e.\ $2\Phi-\theta-\theta_{0} < \pi$.

Finally, the pole $p_{-1}$ describes the wave $\exp[-ik_{0}r\cos(\theta+\theta_0+2\Phi)]$, reflected from the lower face $\theta=-\Phi$ of the wedge. It enters into the result if $2\Phi+\theta+\theta_{0} < \pi$.

Thus, the geometrical optics yields a solution constituted by `pieces' of plane waves. At the ends of these pieces represented by light-shadow boundaries the solution vanishes jumpwise to zero in the shadow region, which means that the geometric-optical solution is discontinuous. These discontinuities of the optical fields will be eliminated with the addition of the diffracted fields in Section \ref{s5}.

The surface wave, grazing towards  the wedge top along the upper face, as shown in Fig.~\ref{Fig:1}, has a complex propagation angle:
\begin{equation}
    \label{4:6}
    \theta_{0}=\Phi-\chi_{+}
    .
\end{equation}
It cannot be reflected from the lower face of the wedge if $\Phi>\tfrac{1}{2}\pi$ as can be readily deduced from the above treatment, and only two poles, namely
\begin{equation}
    \label{4:7}
    p_0 =
    \theta - \Phi + \chi_{+}
    , \qquad
    p_1 =
    \theta  - \Phi - \chi_{+}
    ,
\end{equation}
can fall into the region given by inequation \eqref{4:2}. This occurs if
    \begin{align}
    \label{4:7a}
    \Phi-\pi \mp \Re\chi_{+}
    <\theta < \Phi
    ,
    \end{align}
where the upper and the lower signs stand for $p_{0}$ and $p_{1}$, correspondingly.

Due to the factor $1/\Psi_{0}(\theta_{0})$ in Eq.~\eqref{3:17}, the residue about the pole $p_0$ is evaluated as a wave of unit amplitude:
    \begin{equation}
    \label{4:8}
    B_{0}(r,\theta)  =
    \exp\left[{-ik_{0}r\cos(\theta-\Phi+\chi_{+})}\right]
    .
    \end{equation}
It describes the incident surface wave propagating along the upper face to the top edge of the wedge. According to the inequation \eqref{4:7a}, this wave does not penetrate into the shadow region $\theta < \Phi-\pi$ if $\Re\chi_{+}$ is sufficiently small. We will see in Section \ref{s5} that the transitional region near the formal boundary $\theta=\Phi-\pi-\Re\chi_{+}$ of the shadow region is described by a simple function, which includes the contributions from both the pole $p_{0}$ and the saddle point $p=-\pi$.

The contribution of the pole $p_{1}$ is exactly zero, which can be seen from the fact that formally calculated residue $\Psi_{0}(\theta-p_{1})$ contains the multiplier $\psi_{\Phi}(2\Phi+\pi/2)$, which is zero. Since the contributions of both the $p_{1}$ and $p_{-1}$ poles are zero, the wedge does reflect surface waves (this assertion will be clarified in Section \ref{s6}).


Let us now proceed to the poles of the function $\Psi_{0}(\theta-p)$. To simplify our task, below we restrict ourselves to the case of a right angle wedge with $\Phi=\tfrac{3}{4}\pi$. Then, $\Psi_{0}(\theta-p)$ is expressed through trigonometric functions with the aid of Eq.~\eqref{3:16}, and the poles can be found from the following equation
    \begin{multline}
    \label{4:11}
    \cos\left[
        \frac{1}{6} \left(\theta -\chi_{-}
        -
        p-\frac{\pi }{4}\right)
    \right]
     \cos\left[
        \frac{1}{6}\left(\theta +\chi _{-}
        -
        p-\frac{5\pi}{4}\right)
    \right]
    \\
    \times
    \cos\left[
        \frac{1}{6} \left(\theta -\chi_{+}
        -
        p+\frac{5\pi}{4}\right)
    \right]
    \cos\left[
        \frac{1}{6} \left(\theta +\chi_{+}
        -
        p+\frac{\pi }{4}\right)
    \right]
    =0
    .
    \end{multline}
They obey the conditions of Eq.~\eqref{4:2} if
    \begin{equation}
    \label{4:12}
    \begin{split}
    \tfrac{3}{4}\pi + \Re\chi_{+}
    &<
    \theta \leqslant \phantom{{}-}\tfrac{3}{4}\pi
    ,
    \\
    -\tfrac{3}{4}\pi  \phantom{{} + \Re\chi_{+}}
    &\leqslant \theta
    <  - \tfrac{3}{4}\pi - \Re\chi_{-}
    .
    \end{split}
    \end{equation}
For bare metals, $\Re\chi_{\pm}$ is positive in the optical and infrared ranges of frequencies. Hence, the inequations \eqref{4:12} cannot be satisfied, and the function $\Psi$ cannot have poles inside the contour $g_{0}$. However the zero point
    \begin{equation}
    \label{4:13}
    p_{+} = \theta -\chi_{+}-\tfrac{7}{4}\pi
    \end{equation}
of the factor
    \(
    \cos\left[
        \frac{1}{6} \left(\theta -\chi_{+}
        -
        p+\frac{5\pi}{4}\right)
    \right]
    \)
approaches the first saddle point $p=-\pi$ at $\theta = \Phi = \tfrac{3}{4}\pi$. Similarly, the zero point
    \begin{equation}
    \label{4:15}
    p_{-} = \theta + \chi_{-} + \tfrac{7}{4}\pi
    \end{equation}
of the multiplier
    \(
    \cos\left[
        \frac{1}{6} \left(\theta +\chi_{-}
        -
        p - \frac{5\pi}{4}\right)
    \right]
    \)
is located near the second saddle point $p=+\pi$ at $\theta = - \tfrac{3}{4}\pi$. Although one might expect that these poles strongly affect the diffracted field near the wedge faces at $\theta=\pm\tfrac{3}{4}\pi$, calculations in Section \ref{s6} do not confirm these fears.

\section{Diffracted fields}
\label{s5}

Evaluating the integral in Eq.~\eqref{4:1} along the contours $g_1$ and $g_2$ in Fig.~\ref{Fig:2} yields radiated (freely propagating) electromagnetic fields. Those contours  pass through the saddle points $p=-\pi$ and $p=\pi$, respectively. To use the saddle-point method of integration, we expand the exponent in the integrand of Eq.~\eqref{4:1} about the saddle points so that
    \[
    p
    =
    \mp\pi+\frac{1-i}{\sqrt{k_{0}r}}\,t
    .
    \]
Putting the expansion
\begin{equation*}
    \exp(-ik_{0}r\cos p) \simeq
    \exp[
        ik_{0}r - t^2
    ]
\end{equation*}
in Eq.~\eqref{4:1} reveals that, due to the factor $\exp(-t^2)$, the main contribution to the integral over the contours $g_1$ and $g_2$ comes from the neighborhood of the saddle points. Turning to the integration in the variable $t$ and taking into consideration the direction of integration over the contours $g_1$ and $g_2$, we transform the respective integrals to the following form:
    \begin{subequations}
    \label{5:2}
    \begin{gather}
    \label{5:2a}
    B_{1}(r,\theta)
    =
    \frac{1}{2\pi i}
    \int\limits_{g_{1}}
    s(\theta-p)\,
    \e^{-ik_{0}r\cos p}
    \dif{p}
    =
    -
    \frac{1}{2\pi i}
    \int\limits_{-\infty}^{\infty}
    s\left(\theta + \pi - \frac{1-i}{\sqrt{k_{0}r}}\,t\right)\,
    \e^{ik_{0}r-t^2}
    \der{p}{t}
    \dif{t}
    ,
    \\
    \label{5:2b}
    B_{2}(r,\theta)
    =
    \frac{1}{2\pi i}
    \int\limits_{g_{2}}
    s(\theta-p)\,
    \e^{-ik_{0}r\cos p}
    \dif{p}
    =
    +
    \frac{1}{2\pi i}
    \int\limits_{-\infty}^{\infty}
    s\left(\theta-\pi - \frac{1-i}{\sqrt{k_{0}r}}\,t\right)\,
    \e^{ik_{0}r-t^2}
    \der{p}{t}
    \dif{t}
    .
    \end{gather}
    \end{subequations}
In the wave zone that corresponds to the limit $k_{0}r\to\infty$, the pre-exponential factor $s(\theta-p)$ can be substituted with $s(\theta\pm\pi)$, which yields
    \begin{equation}
    \label{5:4}
    B_{1}(r,\theta) =
    \frac{\e^{ik_{0}r+i\pi/4}}{\sqrt{2\pi k_{0}r}}\,
    s(\theta+\pi)
    ,
    \end{equation}
for the contour $g_{1}$ and
    \begin{equation}
    B_{2}(r,\theta) =
    -
    \frac{\e^{ik_{0}r+i\pi/4}}{\sqrt{2\pi k_{0}r}}\,
    s(\theta-\pi)
    \end{equation}
for the contour $g_{2}$.

 \begin{figure}
  \includegraphics[width=0.8\textwidth]{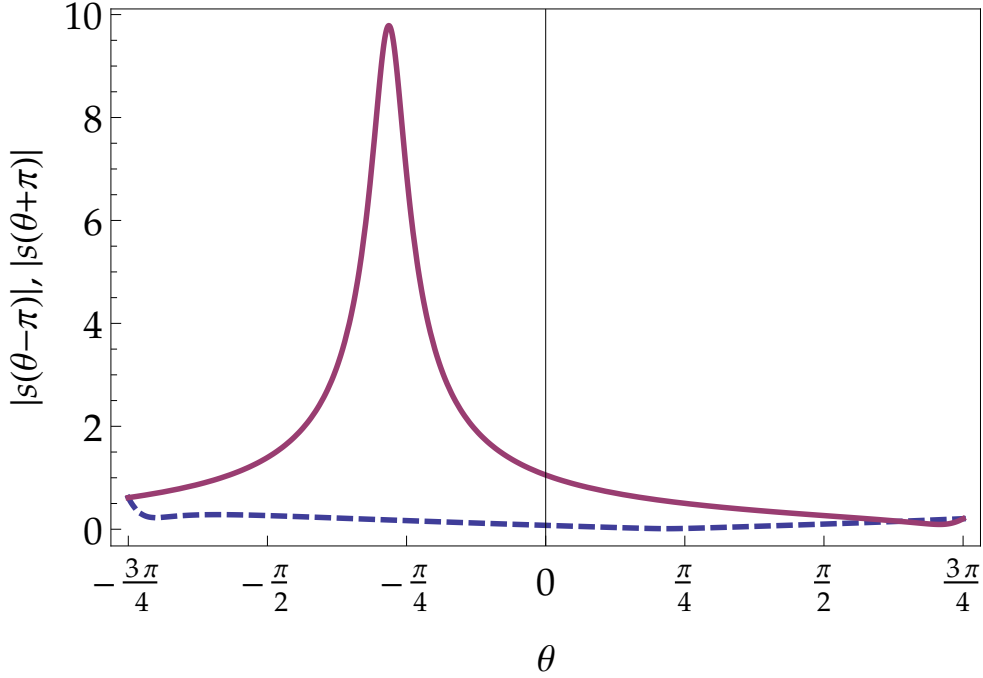}
  \caption{
    (color online)
    Functions $|s(\theta+\pi)|$ (solid line) and $s|(\theta-\pi)|$ (dashed line) for $\chi_{\pm}=0.1(1-i)$.
    The width of the narrow peak at $\theta=-\frac{\pi}{4}$ in the plot of $|s(\theta+\pi)|$ is of the order of $\Im\chi_{+}$. For values of $\chi_{+}$ relevant to the THz range of frequencies, the peak is much narrower and higher.
  }
  \label{Fig:3}
\end{figure}
Since the pole $p_{0}=\theta-\tfrac{3}{4}\pi+\chi_{+}$ approaches the saddle point $p=-\pi$ when $\theta$ tends to $-\tfrac{1}{4}\pi$, the function $s(\theta+\pi)$ has a narrow peak near $\theta=-\tfrac{1}{4}\pi$, which is much bigger than $s(\theta-\pi)$ as shown in Fig.~\ref{Fig:3}. Hence, the field $B_{2}$ can be neglected near the peak, where the amplitude of the diffracted field is
    \begin{equation}
    \label{5:6}
    B_{1} 
    \simeq
    \frac{1}{\sqrt{2\pi k_{0}r}}\,
    \frac{
        \exp\left(
            ik_{0}r+\tfrac{1}{4}i\pi
        \right)
    }{
        \theta +\tfrac{1}{4}\pi + \chi_{+}
    }
    .
    \end{equation}
The angular distribution of the intensity of the radiation field in Eq.~\eqref{5:6} has the Lorentzian profile with the halfwidth
    \begin{equation}
    \label{5:7}
    \Delta\theta = |\Im\chi_{+}|
.
    \end{equation}

\begin{figure}
  \includegraphics[width=\textwidth]{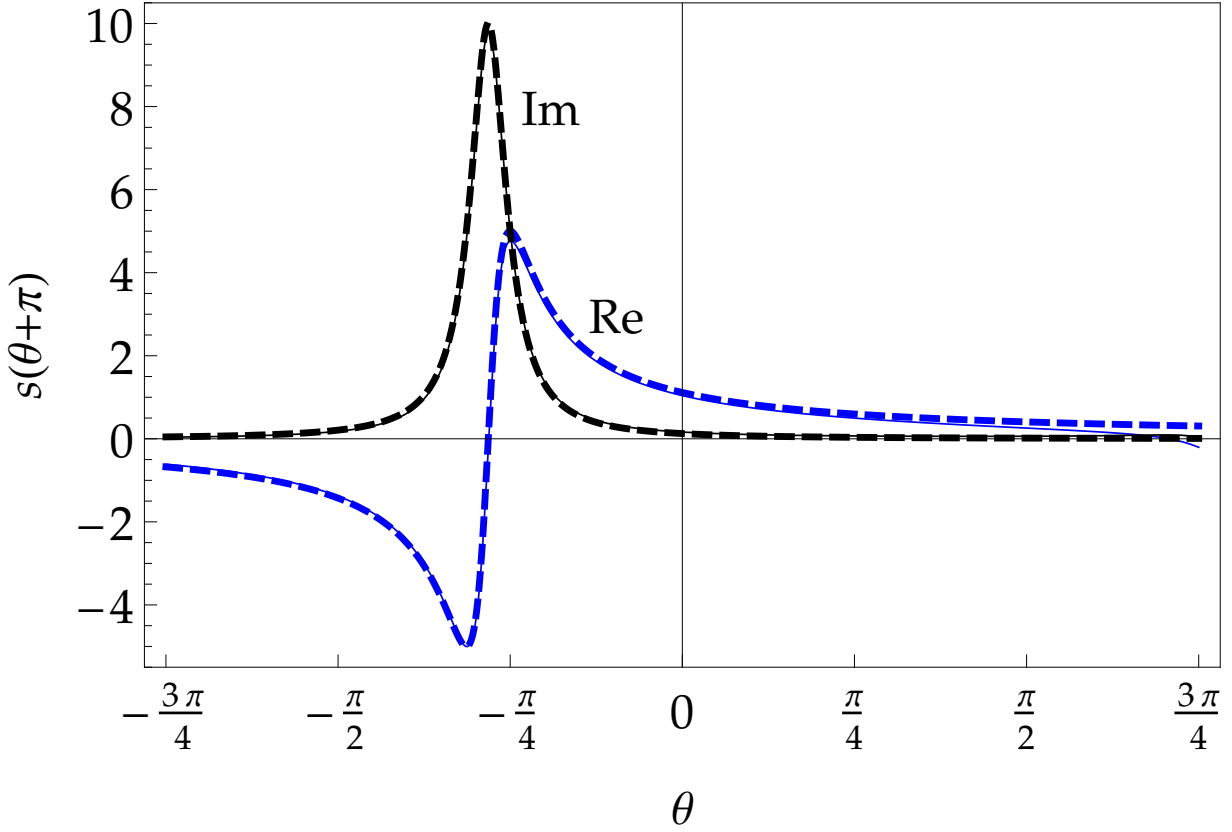}
  \caption{
    (color online)
    Comparison of $s(\theta+\pi)$ (solid line) and Laurent expansion  $s_{0}(\theta+\pi)$ (dashed line); the real part is blue, the imaginary is black;
    $\chi_{+}= 0.1 (1 - i)$.
  }
  \label{Fig:4}
\end{figure}

Similar calculations were done by V.~Zon in \cite{Zon2007JOSA(B)_24_1960,Zon2007JOptA_9_476}.
Note. however. that the range of applicability of Eqs.~\eqref{5:4}–\eqref{5:6} is limited to very large distances, such as
    \begin{equation}
    \label{5:8}
    k_{0} r \gg |\Im\chi_{+}|^{-2} \simeq |\varepsilon|
    .
    \end{equation}
For smaller distances, the factor $s(\theta-p)$ in Eqs.~\eqref{5:2} cannot be considered smooth as compared with the exponent $\exp(-t^2)$ since the former  has a pole at $p_{0} = \theta -\tfrac{3}{4}\pi+\chi_{+}$, which approaches the saddle point $p=-\pi$ at $\theta=-\tfrac{1}{4}\pi$. The absolute magnitudes of the functions $s(\theta\pm\pi)$ are plotted in Fig.~\ref{Fig:3}.
More accurate calculation can be performed using the following approximate expression for the kernel function
    \begin{equation}
    \label{5:9}
    s(\theta-p) \simeq
    s_{0}(\theta-p)
    =
    \frac{1}{\theta - \frac{3}{4}\pi + \chi_{+} - p}
    .
    \end{equation}
It is derived by expanding $s(\theta-p)$ in the Laurent series about the pole $p_0$ but it provides a good approximation for $s(\theta-p)$ in the entire range of integration as Fig.~\ref{Fig:4} proves.
Putting $s_0(\theta-p)$ in Eq.~\eqref{5:2}
yields
    \begin{equation}
    \label{5:10}
    B_{1}(r,\theta)
    =
    \frac{1}{2\pi i}
    \int\limits_{g_{1}}
    s(\theta-p)\,
    \e^{ik_{0}r-t^2}
    \dif{p}
    =
    -
    \frac{\e^{i k_{0} r}}{2\pi i}
    \int_{-\infty}^{\infty}
    \frac{\exp(-t^2)}{z_{0}-t}\dif{t}
    ,
    \end{equation}
where
    \begin{equation*}
    z_{0}
    =
    \frac{\sqrt{k_{0}r}}{1-i}\,(p_{0}+\pi)
    =
    \frac{1+i}{2}
    \left(\theta + \frac{\pi}{4} + \chi_{+}\right)
    \sqrt{k_{0} r}
    .
    \end{equation*}
A method of calculating the integral
    \begin{equation}
    \label{5:12}
    W(z) = \frac{1}{2\pi i} \int _{-\infty}^{\infty}
    \frac{\exp(-t^2)}{z-t}\,\dif{t}
    \end{equation}
in Eq.~\eqref{5:10} is elaborated in the theory of plasma waves \cite{Shafranov1967}. The result is expressed in terms of the imaginary error function
    \[
    \erfi(z) = \frac{2}{\sqrt{\pi}}\int_{0}^{z} \exp\left(x^2\right)\dif{x}
    \]
and has two branches:
    \begin{equation}
    \label{5:14}
    W_{\pm}(z)
    =
    -\frac{1}{2}
    \exp\left(-z^2\right)\left[
        \pm 1
        +
        i \erfi(z)
    \right]
    .
    \end{equation}
The first branch $W_{+}(z)$ is originally computed in the assumption that the imaginary part of $z$ is positive, $\Im(z)>0$, and then analytically continued to the lower half of the complex plane $z$, where $\Im(z)<0$, using Eq.~\eqref{5:14}. The second branch $W_{-}(z)$, on the contrary, is originally computed for $\Im(z)<0$ and then analytically continued to $\Im(z)>0$. In terms of the contour integration in Eq.~\eqref{5:12}, the analytical continuation implies that the pole $t=z$ never crosses the contour of the integration (which originally goes along the axis $\Im(t)=0$), and the contour is deformed when the pole $t=z$ crosses the axis $\Im(t)=0$ to bypass the pole. Since the pole can be bypassed either from above or from below, there appear two branches of the function $W(z)$. They differ in the residue of the integrand at the pole $t=z$:
    \[
    W_{+}(z)-W_{-}(z)
    =
    - \exp(-z^2)
    =
    \frac{1}{2\pi i} \ointctrclockwise\limits_{|t-z|=\delta} \frac{\exp(-t^2)}{z-t}\,\dif{t}
    .
    \]
For large $|z|\gg 1$,  the functions $W_{\pm}(z)$ are evaluated as
    \[
    W_{+} \approx
    \frac{1}{2\sqrt{\pi}i z}
    ,
    \quad
    \text{ and }
    \quad
    W_{-} \approx
    \frac{1}{2\sqrt{\pi}i z} + \exp(-z^2),
    \]
respectively, if $\Im(z)>0$ and as
    \[
    W_{+} \approx
    \frac{1}{2\sqrt{\pi}i z} - \exp(-z^2)
    ,
    \quad
    \text{ and }
    \quad
    W_{-} \approx
    \frac{1}{2\sqrt{\pi}i z}
    ,
    \]
if $\Im(z)<0$. Note that the exponent $\exp(-z^2)$ becomes very big in sectors where $\tfrac{1}{4}\pi < |\arg(z)| < \tfrac{3}{4}\pi$.

The sign of $\Im(z_{0})$ is reversed as the observation point at the angle $\theta$ crosses the boundary $\theta\approx -\tfrac{1}{4}\pi$ of the shadow region. Together with the discontinuous field $B_{0}$, which represents the contribution of Eq.~\eqref{4:8} of the pole $p_{0}$ in the approximation of geometrical optics, the diffracted field $B_{1}$ forms a continuous field. This total continuous field is given by the integral of Eq.~\eqref{5:12} in Eq.~\eqref{5:10} computed for $\Im(z) < 0$. Indeed, the pole $p_{0}$ goes away from the interior of the contour $g_{0}$ through the contour $g_{1}$ if the angle $\theta$ takes a sufficiently large negative value (which corresponds to a deep shadow region) thus making $\Im(z) < 0$. In that case, the integral over the contour $g_{0}$ gives no contribution to the result of calculations so $B_{1}$ stands for the total field. Taking the lower sign in Eq.~\eqref{5:14}, we obtain the final expression
    \begin{equation}
    \label{5:19}
    B(r,\theta)
    =
    -
    W_{-}\left(
        \frac{1+i}{2}
        \left(\theta + \frac{\pi}{4} + \chi_{+}\right)
        \sqrt{k_{0} r}
    \right)
    \e^{ik_{0}r}
    \end{equation}
for the radiation field in the proximity of $\theta=-\tfrac{1}{4}\pi$.
In plasma physics, our choice of the branch $W_{-}(z)$ of the multivalued function $W(z)$ is known as the Landau rule \cite{Shafranov1967}. By analytic continuation, Eq.~\eqref{5:19} represents the total field for any values of $\theta$. It is valid for any $r$ larger than the wavelength in a free space, $r\gg 2\pi/k_0$, both in the near-field and wave zones. The near field was computed numerically in \cite{Zon2007JOSA(B)_24_1960,Zon2007JOptA_9_476} by evaluating the integral in Eq.~\eqref{3:5}. Eq.~\eqref{5:19} gives the desired result with a better accuracy, especially in case where $\Im(\chi_{+})\to 0$.

\begin{figure}
  \includegraphics[width=\textwidth]{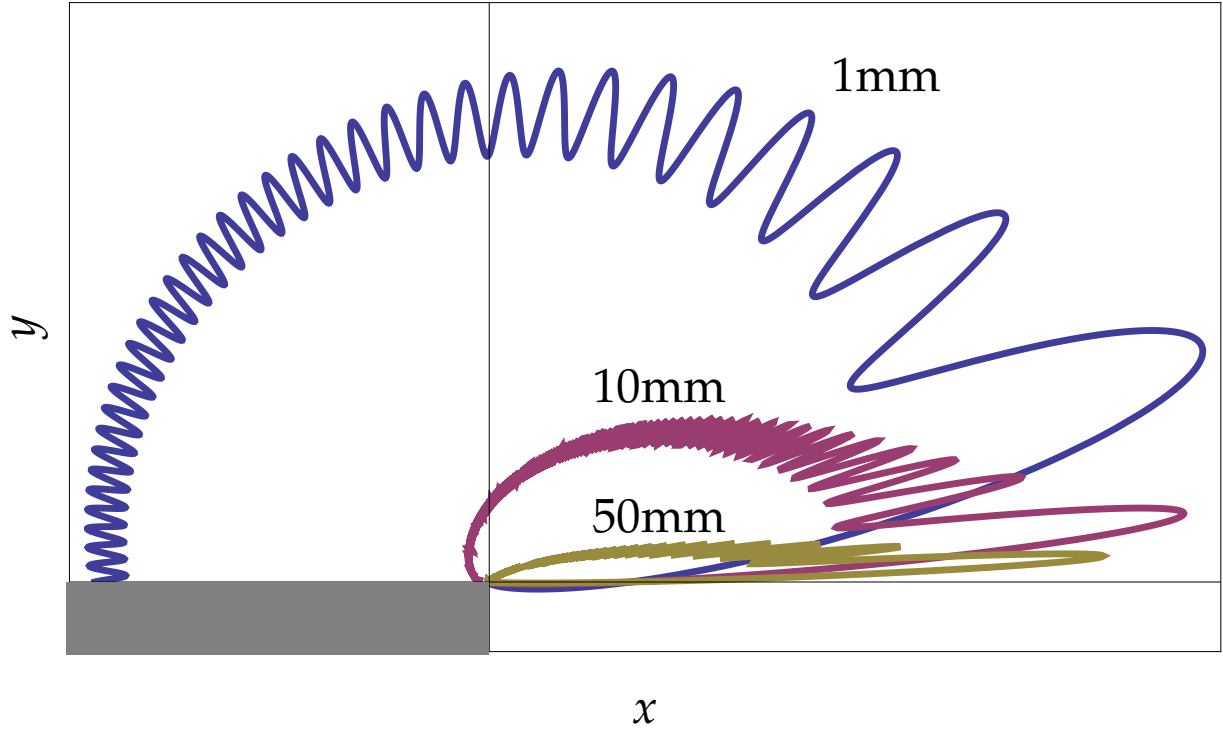}
  \caption{
   (color online)
    An angular pattern of a surface wave diffracted by a pure gold wedge
    for $\lambda=140\,\mu\text{m}$ and $r=1\,\text{mm}$ (blue), $r=10\,\text{mm}$ (magenta), and $r=50\,\text{mm}$ (dark yellow).
    The $x$ and $y$ axes go along the upper and side faces of the wedge as shown in Fig.~\ref{Fig:1}. The distance to the curve is proportional to the intensity of scattered radiation at the given angle.
  }
  \label{Fig:5}
\end{figure}
\begin{figure}
  \includegraphics[width=\textwidth]{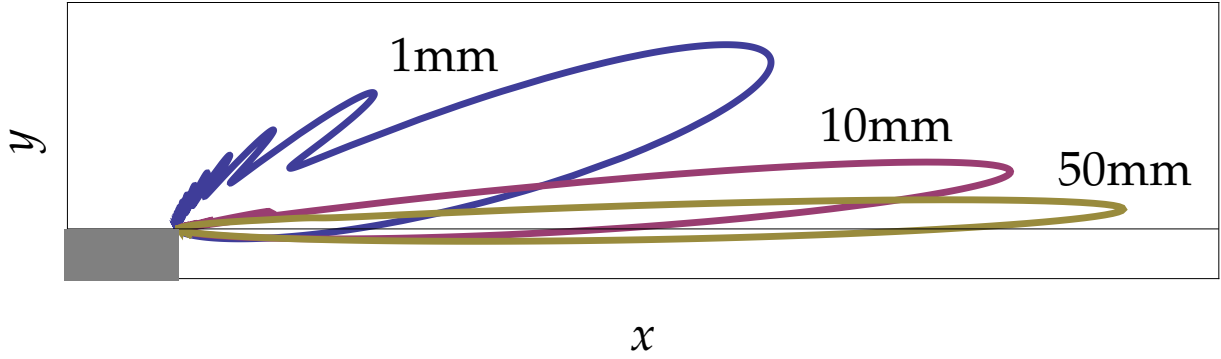}
  \caption{
    (color online)
    Same as in Fig.~\ref{Fig:5} for a gold wedge coated with $0.75\mu\text{m}$ ZnS film:
    $r=1\,\text{mm}$ (blue), $r=10\,\text{mm}$ (magenta, scaled $\times4$), $r=50\,\text{mm}$ and (dark yellow, scaled $\times16$).
  }\label{Fig:6}
\end{figure}

Angular patterns of diffracted waves computed by formula in Eq.~\eqref{5:19} for a pure gold substrate and a gold wedge coated with ZnS film are presented in Figs.~\ref{Fig:5} and~\ref{Fig:6}, respectively, where the intensity $|B|^2$ is depicted for several distances from the edge of the wedge for parameters relevant to ongoing experiments at Novosibirsk Free Electron Laser. Fig.~\ref{Fig:5} is typical for the near-field zone, where the angular distribution has many peaks. Fig.~\ref{Fig:6}, on the contrary, shows angular distributions with a single peak characterized by the Lorentzian profile of Eq.~\eqref{5:6}. In the first case of a pure gold wedge the far wave zone begins after the distance $r\approx 15\,\text{m}$ from the wedge edge whereas for ZnS-coated gold the wave zone is located at $r\approx 20\,\text{mm}$.

In contrast to the Lorentzian profile Eq.~\eqref{5:6}, the total diffracted field in Eq.~\eqref{5:19} is not symmetric about the boundary of the shadow region $\theta\approx-\tfrac{1}{4}\pi$. It extends into the illuminated region $\theta>-\tfrac{1}{4}\pi$ much further than into the shadow region $\theta<-\tfrac{1}{4}\pi$. This assertion has found a convincing evidence in experimental data cited in \cite{Gerasimov+2012b}.
Another important conclusion is that the radiation intensity in the illuminated area does not copy the profile of the surface wave: instead of monotonic decreasing with the distance from the face of the wedge it rises at first. Again, this fact has been confirmed experimentally \cite{Gerasimov+2012b}.



\begin{figure}
  \includegraphics[width=\textwidth]{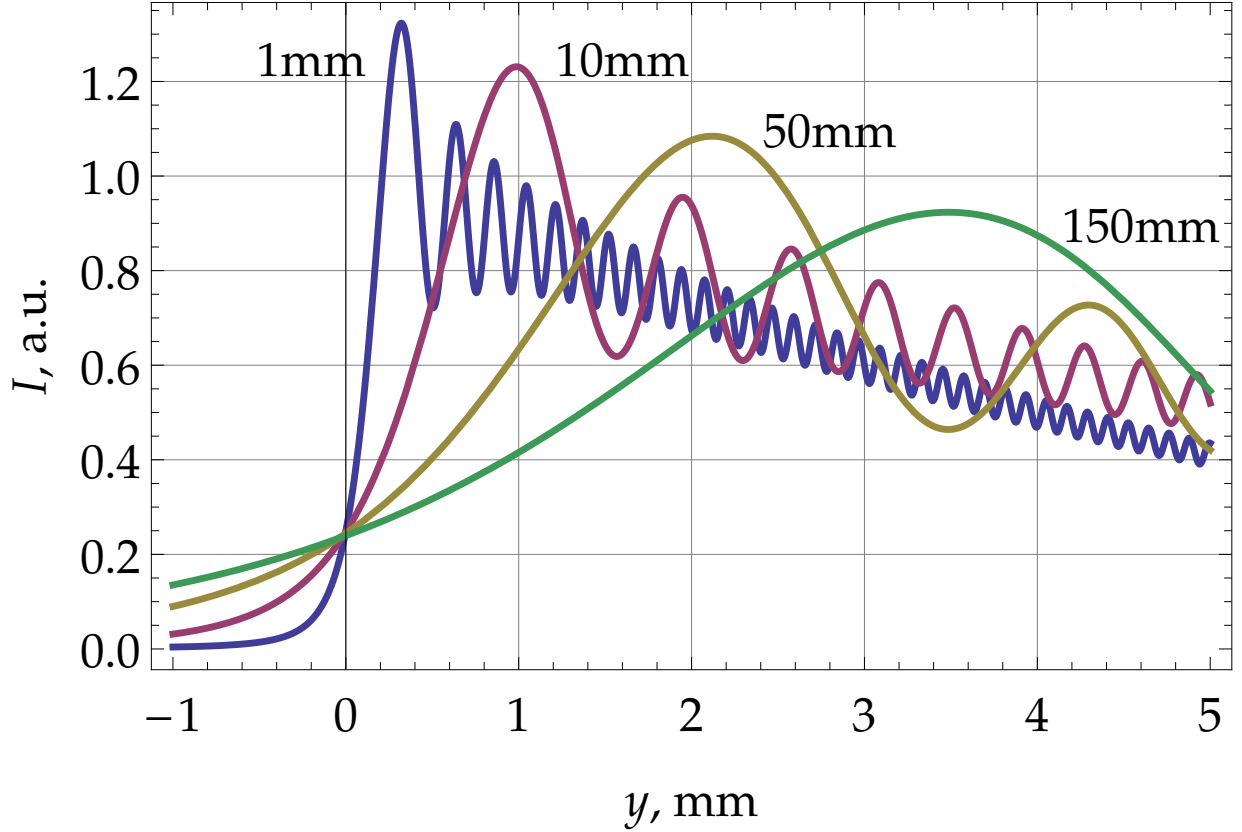}
  \caption{
    (color online)
    Intensity of diffracted wave vs.\ vertical coordinate at different distances from the wedge edge: $x=1\,\text{mm}$ (blue), $x=10\,\text{mm}$  (magenta), $x=50\,\text{mm}$ (dark yellow), and
    $x=150\,\text{mm}$ (green).
    Pure gold wedge.
  }\label{Fig:7}
\end{figure}
\begin{figure}
  \includegraphics[width=\textwidth]{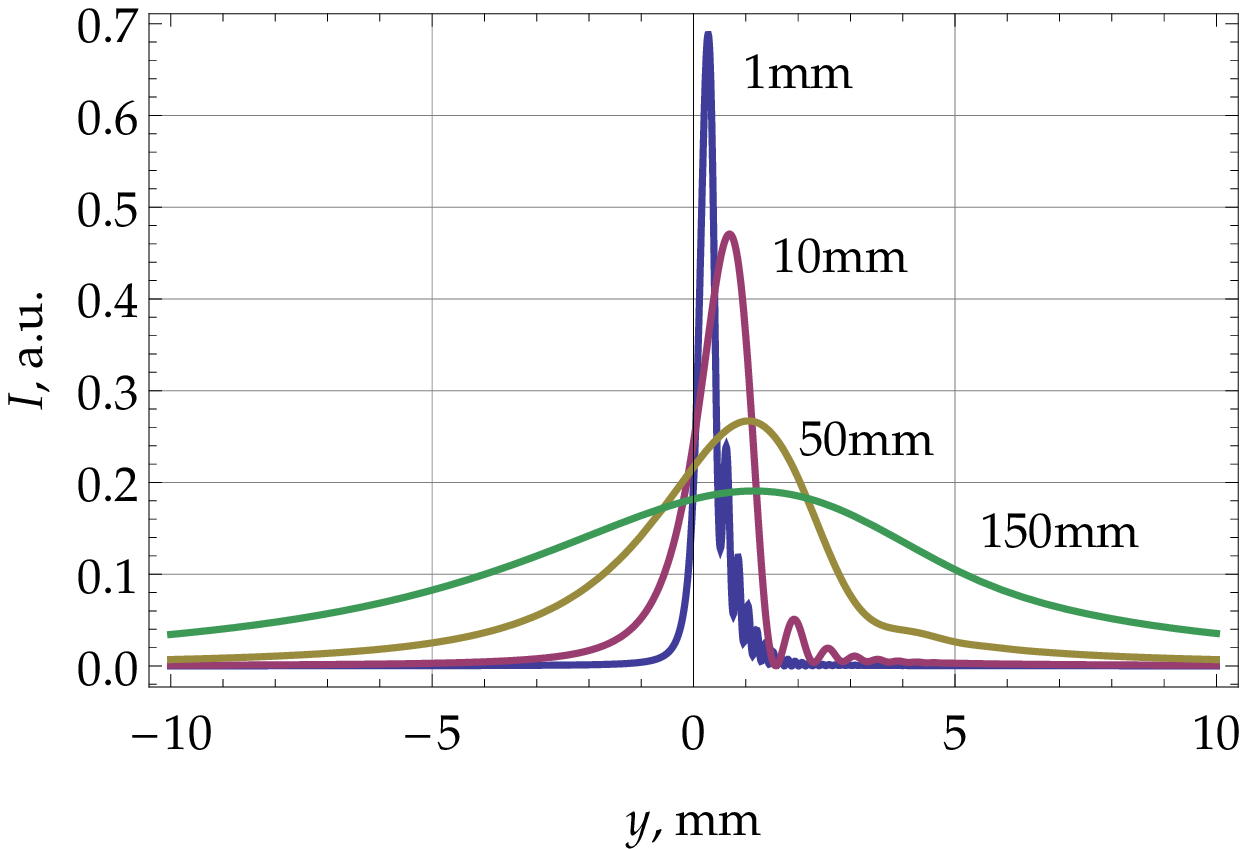}
  \caption{
    (color online)
    Intensity of diffracted wave vs.\ vertical coordinate at different distances from the wedge edge: $x=1\,\text{mm}$ (blue), $x=10\,\text{mm}$  (magenta, scaled $\times2$), $x=50\,\text{mm}$ (dark yellow, scaled $\times4$), and , $x=150\,\text{mm}$ (green, scaled $\times8$).
    Gold wedge coated with a $0.75\mu\text{m}$ ZnS film.
  }\label{Fig:8}
\end{figure}
Intensity of diffracted waves vs.\ distance from the upper face is shown in Figs.~\ref{Fig:7} and~\ref{Fig:8}, which illustrate that incident surface wave is mainly scattered into the upper hemisphere.

\section{Scattered fields near wedge faces}
\label{s6}

Near the face surfaces of the wedge at $\theta=\pm\tfrac{3}{4}\pi$ the integrals $B_{1}$ and $B_{2}$ have commensurate magnitude and both should be computed. Moreover, the integrands in Eqs.~\eqref{5:2} obey the equality
    \begin{equation}
    \label{6:1}
    s(\theta+\pi)=s(\theta-\pi)
    \end{equation}
at $\theta=\pm\tfrac{3}{4}\pi$ as can be seen from Fig.~\ref{Fig:3}. The approximation of Eq.~\eqref{5:9} provides a good fit for the kernel $s(\theta-p)$ near the saddle point $p=-\pi$ for almost entire interval $-\tfrac{3}{4}\pi < \theta < \tfrac{3}{4}\pi$ of the angles $\theta$. However it does not satisfy Eq.~\eqref{6:1} at the ends and could be improved there.

Near the upper face at $\theta=\tfrac{3}{4}\pi$ the pole $p_{+}=\theta -\chi_{+}-\tfrac{7}{4}\pi$ approaches the saddle point $p=-\pi$ at $\theta=\tfrac{3}{4}\pi$, as mentioned at the end of Section \ref{s4}.
Expansion of the kernel $s(\theta-p)$ in Eq.~\eqref{5:2a} into the Laurent series about the pole $p_{+}=\theta -\frac{7}{4}\pi - \chi_{+}$ near the saddle point $p=-\pi$
has the following form:
    \begin{equation}
    \label{6:2}
    s(\theta-p) \simeq
    \frac{1}{3\sqrt{3}}
    \left(
        1 + \frac{2 \chi_{+} }{\theta - \frac{7}{4}\pi - \chi_{+} - p}
    \right)
    .
    \end{equation}
The kernel
    \begin{equation}
    \label{6:3}
    s(\theta-p) \simeq
    -\frac{1}{3\sqrt{3}}
    \end{equation}
in Eq.~\eqref{5:2b} is regular at $\theta=\tfrac{3}{4}\pi$ near the second saddle point $p=\pi$. Combining Eqs.~\eqref{5:2}, \eqref{6:2}, and \eqref{6:3}, we obtain
    \begin{multline}
    \label{6:4}
    B_{1}+B_{2}
    \simeq
    -\frac{1}{2\pi i} \int_{-\infty}^{\infty}
    \frac{2}{3\sqrt{3}}
    \left(
        1 + \frac{\chi_{+} }{\theta - \frac{3}{4}\pi - \chi_{+} - \der{p}{t} t}
    \right)
    \e^{ik_{0}r-t^2}\,\der{p}{t}\,\dif{t}
    \\
    =
    \frac{2\e^{ik_{0}r}}{3\sqrt{3}}\,
    \left[
    \frac{\e^{i\pi/4}}{\sqrt{2\pi k_{0}r}}
    -
    \chi_{+}
    W_{-}\left(
        \frac{1+i}{2}
        \left(\theta-\frac{3}{4}\pi - \chi_{+}\right)
        \sqrt{k_{0}r}
    \right)
    \right]
    .
    \end{multline}
The second term in Eq.~\eqref{6:4}, containing $\chi_{+}W_{-}$, is small and could be neglected. Eq.~\eqref{5:19} gives a similar result at $\theta=\tfrac{3}{4}\pi$, which differs by a numerical coefficient of the order of unity.

Similarly, the pole $p_{-}=\theta +\frac{7}{4}\pi + \chi_{-}$ approaches the saddle point $p=\pi$ at $\theta\to -\tfrac{3}{4}\pi$. The expansion of the kernel of the integral in Eq.~\eqref{5:2b} about $p_{-}$ has the following form:
    \begin{equation}
    \label{6:6}
    s(\theta-p) \simeq
    s_{-}(\theta-p)
    =
    \frac{1}{\sqrt{3}}
    \left(
        1
        -
        \frac{2\chi_{-}}{\theta+\frac{7}{4}\pi+\chi_{-}-p}
    \right)
    ,
    \end{equation}
and the kernel of Eq.~\eqref{5:2a} is
    \begin{equation}
    \label{6:7}
    s(\theta-p) \simeq
    -\frac{1}{\sqrt{3}}
    .
    \end{equation}
Combining Eq.~\eqref{5:2} with Eqs.~\eqref{6:6} and~\eqref{6:7} yields
    \begin{multline}
    \label{6:8}
    B_{1}+B_{2}
    \simeq
    \frac{1}{2\pi i} \int_{-\infty}^{\infty}
    \frac{2}{\sqrt{3}}
    \left(
        1 - \frac{\chi_{-} }{\theta + \frac{3}{4}\pi + \chi_{-} - \der{p}{t} t}
    \right)
    \e^{ik_{0}r-t^2}\,\der{p}{t}\,\dif{t}
    \\
    =
    -
    \frac{2\e^{ik_{0}r}}{\sqrt{3}}\,
    \left[
    \frac{\e^{i\pi/4}}{\sqrt{2\pi k_{0}r}}
    +
    \chi_{-}
    W_{+}\left(
        \frac{1+i}{2}
        \left(\theta+\frac{3}{4}\pi + \chi_{-}\right)
        \sqrt{k_{0}r}
    \right)
    \right]
    .
    \end{multline}
Again, the second term with $\chi_{-}$ is small and can be dropped.

Eq.~\eqref{6:4} describes a wave propagating along the upper face of the wedge back from the wedge edge, and Eq.~\eqref{6:8} is a wave propagating along the lower face. They can be thought of as reflected and refracted waves, respectively, although they are not surface waves. It is interesting that the amplitude of the refracted wave is $3$ times larger than the amplitude of the reflected one.

\section{Comparison with experiment}
\label{s7}

\begin{figure}
  \centering
  \includegraphics[width=\textwidth]{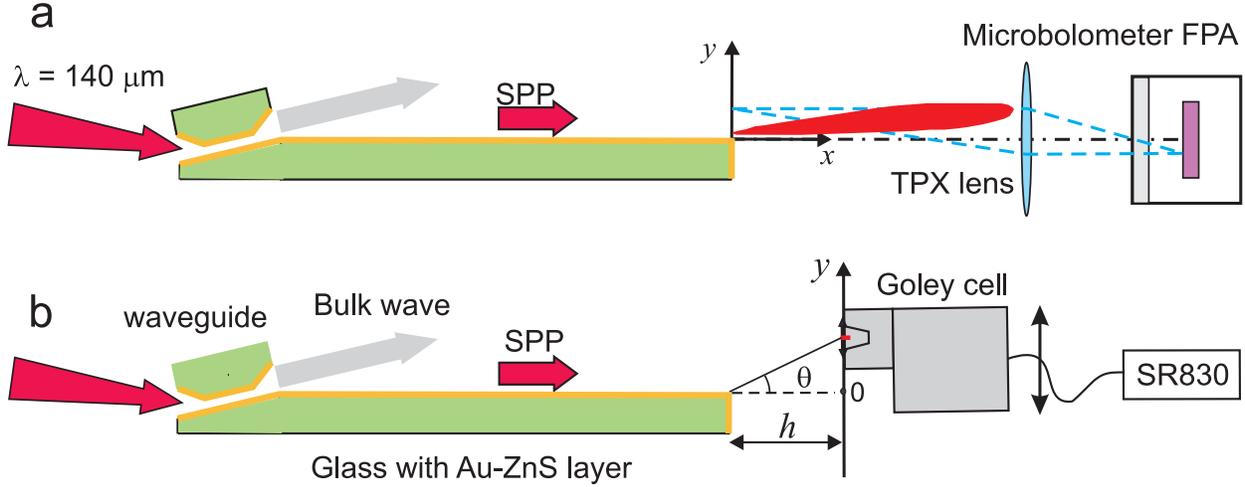}
  \caption{
    (color online)
    Experimental configurations for the study of radiation induced by decoupled SPP using (a) a microbolometer focal plane array and (b) a Goley cell coupled to a lock-in amplifier SR830.
  }\label{Fig:9}
\end{figure}
The experimental configurations are shown in Fig.~\ref{Fig:9}. Monochromatic radiation of Novosibirsk Free Electron Laser (FEL) at a frequency of $2.3\,\text{THz}$ ($\lambda = 140\,{\mu}\text{m}$) entered the user station through a 16-m long beamline as a Gaussian beam $I=I_{0}\exp\left(-2r^2/w^2\right)$  with a waist of $9\,\text{mm}$. After passing a circular aperture with a diameter of $10\,\text{mm}$, the beam was focused with a cylindrical mirror into the input mouth of a plane waveguide, formed by the gold-covered facets of two glass prisms. At the output mouth of the waveguide, the radiation passed through the slit, transformed into a surface plasmon-polariton travelling along the metal-dielectric interface. Since a portion of the radiation could be emitted as a free wave, to separate the SPP and the bulk wave we made the input facet of the large prism tapered with an angle of 13 degrees to the upper sample plane. The latter was $17\,\text{cm}$ long and $4\,\text{cm}$ wide. The edge between these two facets was smoothed for the purpose of decreasing the SPP radiation loss.

The large and small facets of glass slabs were covered with a $1\,\mu\text{m}$ thick gold layer, which was considered within this problem as a bulk metal since the skin-depth for the gold is much smaller than $1\,\mu\text{m}$. The SPPs, which were launched with the help of the waveguide, travelled along the large facet of the samples. Bare gold and gold covered with ZnS layers $0.1$ to $3$ $\mu\text{m}$ thick were employed in the experiments. To study SPP characteristics we applied two non-invasive techniques for detection of electromagnetic (EM) radiation in the free space behind the end facet of the samples. An optical system consisting of a TPX lens with $f=50\,\text{mm}$ \cite{tydexoptics.com} and a microbolometer focal plane array \cite{Dem'yanenko+2008AplPhysLett_92_131116, Knyazev+JIMTW_32_1207} was used for imaging of the EM radiation wavefronts. Intensity of the EM radiation at different distances $h$ was scanned along the $y$-axis using an opto-acustic Goley cell \cite{tydexoptics.com} with an input slit $0.2\,\text{mm}$ thick directed along the $z$-axis.

\begin{figure}[!ptb]
  \centering
  \includegraphics[width=\textwidth]{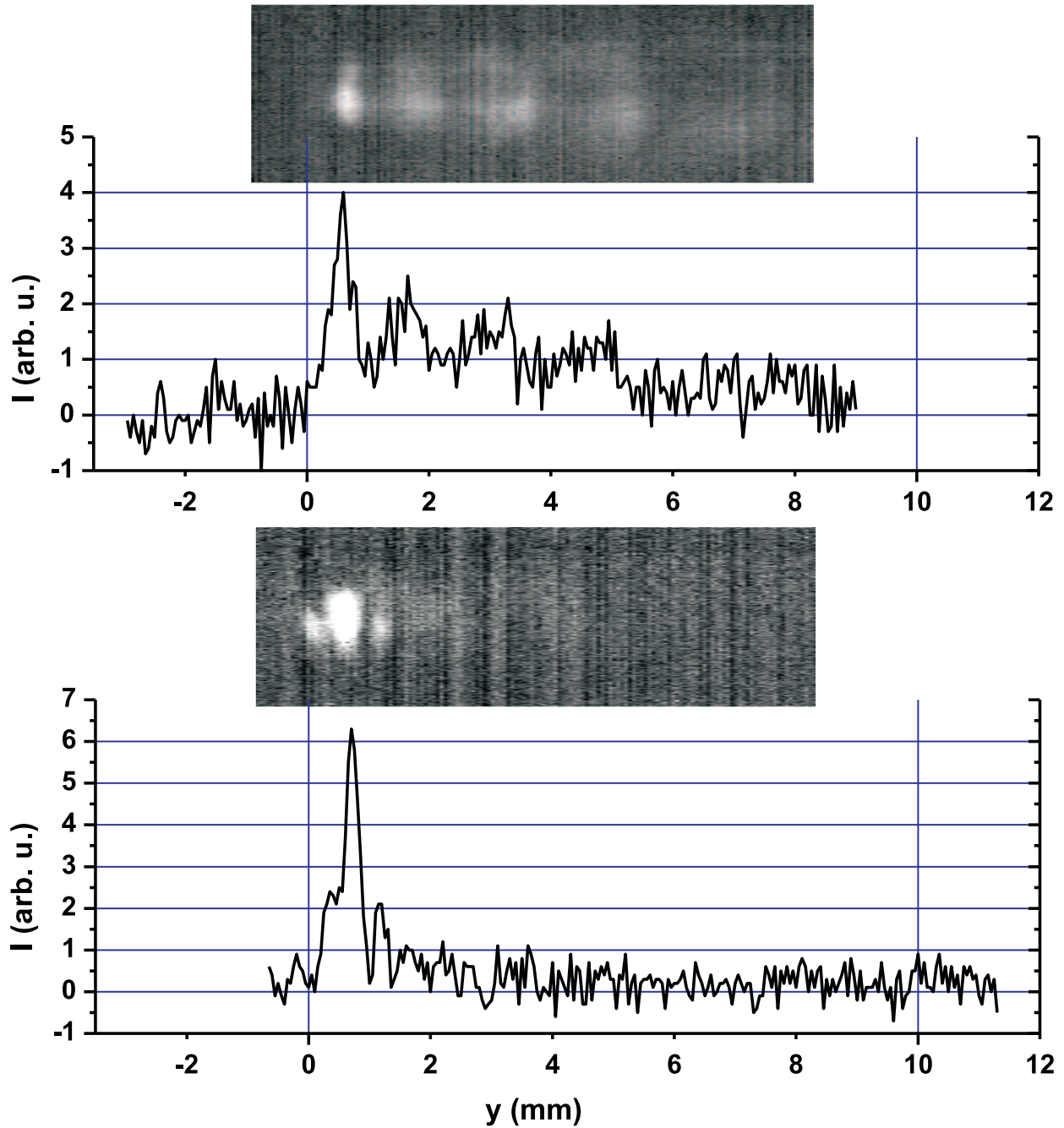}
  \caption{
    Images of spatial distribution of radiation near the end of a bare gold sample (above) and a sample covered with a 1-${\mu\text{m}}$ ZnS layer (below), recorded with the microbolometer focal plane array (configuration of Fig.~\ref{Fig:9},a).
  }\label{Fig:10}
\end{figure}
\begin{figure}[!ptb]
  \centering
  \includegraphics[width=\textwidth]{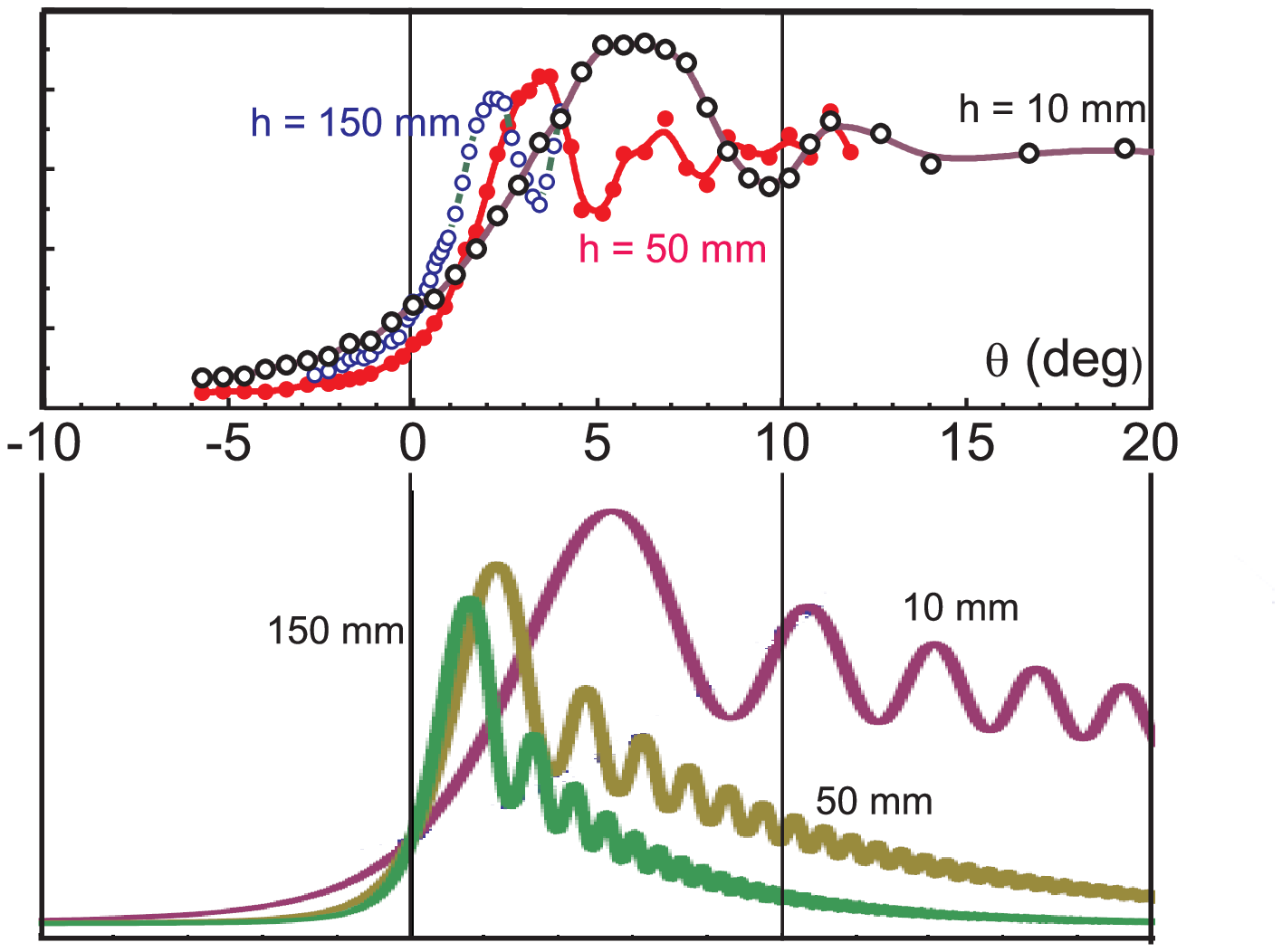}
  \caption{
    EM wave intensity distribution (in
    arbitrary units) vs. angle for three distances in the experimental configuration of Fig.~\ref{Fig:9}, b for a bare gold surface: (a) Goley cell signal for three distances from the sample end; (b) simulations.
  }\label{Fig:11}
\end{figure}
The experimental results were presented in brief in the conference proceedings \cite{Gerasimov+2012b} and will be published in detail elsewhere \cite{Gerasimov+2013}. In this paper we present only the data relevant to the contents of the theory, namely we describe characteristics of the electromagnetic field (EMF) that arose when an SPP reached the end of the surface. The main features of the EMF were as follows (see Figs.~\ref{Fig:10} and~\ref{Fig:11}). (a) The intensity of EMF, in contrast to the SPP intensity, reached its maximum at a distance of 1 to 2 mm above the sample surface, depending on the ZnS layer thickness, and only then sloped down. (b) Oscillations were observed on the descending part of the distribution. (c) The width of the distribution decreased with the distance and the angular distribution became narrower. All these features are in reasonable agreement with the theoretical predictions. Some discrepancies between the distribution shapes on the slope recorded with the MBFPA and the Goley cell may be caused, for example,  by field disturbance by the metal diaphragm at the cell input window.

\section{Conclusions}
\label{s9}

In this paper, we have considered diffraction of a surface wave by a rectangular wedge with impedance facets using the Sommerfeld-Malyuzhinets technique. We have derived Eq.~\eqref{5:19} that uniformly describes the diffracted field  both in the near-field and far-field zones. It was used in Section \ref{s5} for analysis of diffracted radiation at various distances from the wedge. Main conclusions from this formula are confirmed by available data from ongoing experiments at Novosibirsk Free Electron Laser \cite{Gerasimov+2012b,Gerasimov+2013}.

First, we have shown that the total diffracted field expressed by Eq.~\eqref{5:19} is not symmetric about the boundary of the shadow region and that the surface wave is scattered mainly into the upper hemisphere. It extends into the illuminated region  much further than into the shadow region as depicted in Figs.~\ref{Fig:7} and \ref{Fig:8}.

Another important conclusion is that the radiation intensity in the illuminated area does not follow a surface wave profile and at first increases instead of monotonic decreasing with the distance from the face of the wedge, as one might expect for the surface wave, which exponentially decays with the distance from the metal-air interface. Again, this fact has been confirmed experimentally in \cite{Gerasimov+2012b,Gerasimov+2013}.

We have confirmed the conclusion made in \cite{Zon2007JOSA(B)_24_1960, Zon2007JOptA_9_476} that in the wave zone the angular distribution of the scattered wave has the Lorentzian form with a width determined by impedance. However, we have noted that the wave zone for a wedge with a small surface impedance begins at very large distances, specified by Eq.~\eqref{5:8}.

\section*{Acknowledgements}


This work was supported by the Russian Government, grant 11.G34.31.0033, the Russian Foundation for Basic Research, grants 11-02-12252-ofi-m, 11-02-12171-ofi-m , and the Russian Ministry of Education and Science, state contracts No 14.B37.21.0732, 14.B37.21.0784 and 14.B37.21.0750. The experiments mentioned in the paper were carried out using equipment belonging to the SCSTR.

\end{document}